\documentclass[twocolumn,preprintnumbers,amsmath,amssymb]{revtex4}

\usepackage{mathrsfs}%====================================
\usepackage{amssymb}
\usepackage{graphicx}% Include figure files
\usepackage{dcolumn}% Align table columns on decimal point
\usepackage{bm}% bold math
\usepackage{textcomp}
\usepackage{amsmath}
\usepackage[titletoc]{appendix}

\newcommand\ket[1]{\ensuremath{|#1\rangle}}

\newcommand\oprod[2]{\ensuremath{|#1\rangle\langle#2|}}
\newcommand\mean[1]{\ensuremath{\langle #1\rangle}}

\newcounter{RomanNumber}

\begin{document}
\title{Unconditional security of sending or not sending twin-field quantum key distribution with finite pulses}
\author{ Cong Jiang$ ^{1,2}$, Zong-Wen Yu$ ^{1,3}$, Xiao-Long Hu$ ^{1,2}$
and Xiang-Bin Wang$ ^{1,2,4\footnote{Email Address: xbwang@mail.tsinghua.edu.cn}\footnote{Also at Center for Atomic and Molecular Nanosciences, Tsinghua University, Beijing 100084, China}}$}

\affiliation{ \centerline{$^{1}$State Key Laboratory of Low
Dimensional Quantum Physics, Department of Physics,} \centerline{Tsinghua University, Beijing 100084,
Peoples Republic of China}
\centerline{$^{2}$ Synergetic Innovation Center of Quantum Information and Quantum Physics, University of Science and Technology of China}\centerline{  Hefei, Anhui 230026, China
 }
\centerline{$^{3}$Data Communication Science and Technology Research Institute, Beijing 100191, China}\centerline{$^{4}$ Jinan Institute of Quantum technology, SAICT, Jinan 250101,
Peoples Republic of China}}
%%%%%%%%%%%%%%%%%%%%%%%%%%%%%%%%%%%%%%%%%%%%%%%%%%%%%%%%%%%%%%%%%%%
%%%%%%%%%%%%%%%%%%%%%%%%%%%%%%%%%%%%%%%%%%%%%%%%%%%%%%%%%%%%%%%%%%%
%%%%%%%%%%%%%%%%%%%%%%%%% Abstract %%%%%%%%%%%%%%%%%%%%%%%%%%%%%%%%
\begin{abstract}
The Sending-or-Not-Sending protocol of the twin-field quantum key distribution (TF-QKD) has its advantage of unconditional security proof under any coherent attack and fault tolerance to large misalignment error. So far this is the only coherent-state based TF-QKD protocol that has considered finite-key effect, the statistical fluctuations. Here we consider the complete finite-key effects for the protocol and we show by numerical simulation that the protocol with typical finite number of pulses in practice can produce unconditional secure final key under general attack, including all coherent attacks. It can exceed the secure distance of 500 $km$ in typical finite number of pulses in practice even with a large misalignment error. 
\end{abstract}

%%%%%%%%%%%%%%%%%%%%%%%%%%%%%%%%%%%%%%%%%%%%%%%%%%%%%%%%%%%%%%%%%%%
%%%%%%%%%%%%%%%%%%%%%%%%%%%%%%%%%%%%%%%%%%%%%%%%%%%%%%%%%%%%%%%%%%%
%%%%%%%%%%%%%%%%%%%%%%%%%%%%%%%%%%%%%%%%%%%%%%%%%%%%%%%%%%%%%%%%%%%

\maketitle
\section{Introduction}
Quantum key distribution (QKD) could provide unconditionally secure communication~\cite{bennett1984quantum,gisin2002quantum,gisin2007quantum,scarani2009security,shor2000simple,koashi2009simple,tamaki2003unconditionally,kraus2005lower} of two parties, Alice and Bob. But the security in ideal case~\cite{shor2000simple,koashi2009simple,tamaki2003unconditionally,kraus2005lower} dose not guarantee the security in practice~\cite{huttner1995quantum,yuen1996quantum,brassard2000limitations,lu2000security,lu2002quantum,lydersen2010hacking,
gerhardt2011full,hayashi2007upper,scarani2008quantum}. Fortunately, the decoy-state method~\cite{hwang2003quantum,wang2005beating,lo2005decoy,wang2007quantum,rosenberg2007long,schmitt2007experimental,
peng2007experimental,liao2017satellite,peev2009secoqc,chen2010metropolitan,sasaki2011field,frohlich2013quantum,
boaron2018secure,wang2008experimental,wang2005decoy,ad2007simple,wang2007simple,wang2008general,wang2009decoy,
tamaki2014loss,yu2016reexamination,xu2009experimental,chau2018decoy} could help us beating the photon-number-splitting (PNS) attack~\cite{huttner1995quantum,yuen1996quantum,brassard2000limitations,lu2000security,lu2002quantum} and guarantee the security with imperfect light sources. Besides decoy-state mehtod, there are other protocols such as RRDPS protocol~\cite{sasaki2014practical,takesue2015experimental} proposed to beat PNS attack.  Measurement-device-independent (MDI)-QKD~\cite{braunstein2012side,lo2012measurement} can solve all possible loopholes of detection. And the decoy-state MDI-QKD \cite{
wang2013three,rubenok2013real,liu2013experimental,tang2014experimental,tang2014measurement,wang2015phase,comandar2016quantum,
yin2016measurement,wang2017measurement,curty2014finite,xu2014protocol,
yu2015statistical,zhou2016making,jiang2017measurement} protocol could help us ensure the security of protocol performed by imperfect light sources and detectors.

The 4-intensity protocol~\cite{zhou2016making} together with the joint-constraints~\cite{yu2015statistical} has greatly improved the key rate and distance of the MDI-QKD. Using this protocol, a distance exceeding 400 km has been experimentally demonstrated~\cite{yin2016measurement} for the MDI-QKD. However, the key rate of all the prior art decoy-state protocols and the MDI-QKD protocols protocols cannot be better than the linear scale of the channel transmittance. It cannot exceed the known bound of the repeaterless QKD, such as the PLOB bound~\cite{pirandola2017fundamental} or the TGW bound~\cite{takeoka2014fundamental}. Recently, a QKD protocol named Twin-Field (TF) QKD was proposed~\cite{lu2018overcoming} whose key rate $R\sim O(\sqrt{\eta})$, where $\eta$ is the channel transmittance, and thus has attracted much attention. But the later announcement of the phase information in Ref.~\cite{lu2018overcoming} will casuse security loopholes~\cite{wang2018effective,wang2018twin}, and many variants of TF-QKD have been proposed~\cite{wang2018twin,tamaki2018information,ma2018phase,cui2019twin,curty2018simple,yu2019sending,lu2019twin} to close the loophole. A series of experiments~\cite{minder2019experimental,liu2019experimental,wang2019beating,zhong2019proof} have been done to demonstrate those protocols. In particular, an efficient protocol for TF-QKD through sending-or-not-sending (SNS protocol) has been given in Ref.~\cite{wang2018twin}. The SNS protocol has been experimentally demonstrated in proof-of-principle in Ref.~\cite{minder2019experimental}, and realized in real optical fiber with the effects of statistical fluctuation being taken~\cite{liu2019experimental}. The unconditional security of SNS protocol in the asymptotic case has been proved~\cite{wang2018twin} and SNS protocol relaxes the requirement for single photon interference accuracy. The key rate of SNS is still considerable even if the misalignment error is as large as $35\%$. Among all those variants of TF QKD with coherent states, the SNS QKD protocol is the only one that takes the effect of statistical fluctuation and finite decoy states into consideration~\cite{yu2019sending}. Here we show an analysis of the complete effect of finite-key size of SNS QKD protocol.

The main tool we use to analyse the effect of finite-key size is the universally composable framework~\cite{muller2009composability}. An complete QKD protocol usually includes the preparation and distribution of quantum states, measurement of received quantum states, parameter estimation, error correction and private amplification. After the error correction step, Alice gets a bit string $S$, and Bob get an estimate string $S^\prime$ of $S$. If the error rate is too large, the results of error correction is an empty string and the protocol aborts. A protocol is called $\varepsilon_{cor}$-correct if the probability that $S$ and $S^\prime$ aren't the same, Pr$(S \neq S^\prime)\le \varepsilon_{cor}$.

Besides, the quantum state of Alice may be attacked by Eve in the distribution and measurement steps and some information would be leaked to Eve. To ensure the security of final secret keys, Alice and Bob apply a privacy amplification scheme based on two-universal hashing~\cite{renner2005security} to extract two shorter strings of length $l$ from $S$ and $S^\prime$. We denote the density operator of the system of Alice and Eve as $\rho_{AE}$. If 
\begin{equation}
\min_{\rho_E}\frac{1}{2}\parallel \rho_{AE}-U_A\otimes\rho_E\parallel\le \varepsilon_{sec},
\end{equation}
where $U_A$ denotes the fully mixed state of Alice's system of strings of length $l$ and $\rho_E$ is the density operator of Eve's system, the protocol is called $\varepsilon_{sec}$-secret~\cite{curty2014finite,konig2007small,tomamichel2012tight}. According to the composable framework, a protocol is called $\varepsilon$-secure if it is both $\varepsilon_{cor}$-correct and $\varepsilon_{sec}$-secret, and $\varepsilon\le \varepsilon_{cor}+\varepsilon_{sec}$.

This paper is arranged as follows. In Sec.~\ref{sec2}, we introduce the main results of the effect of finite-key size. And in Sec.~\ref{simulation}, we present our numerical simulation results. The article ends with some concluding remarks. The details of calculation are shown in the Methods part.

\section{The effect of finite-key size of SNS protocol}\label{sec2}
As shown in Ref.~\cite{wang2018twin}, there are two windows in SNS protocol, the $\widetilde{X}$ windows and the $Z$ windows. In a $Z$ window, Alice (Bob) randomly decides to send a phase-radomized coherent state $\ket{\sqrt{\mu_z}e^{i\theta_A}}$ ($\ket{\sqrt{\mu_z}e^{i\theta_B}}$) with a probability $p$, or sends nothing (a vacuum state \ket{0}). In an $\widetilde{X}$ window (note that the $\widetilde{X}$ window defined here is a slightly different from the definition of the $X$ window in Ref.~\cite{wang2018twin}, thus we use a different symbol.), Alice and Bob randomly send out a phase-randomized coherent state.

The $\widetilde{X}$ windows are decoy windows and will be used to estimate the counting rate $s_1$ and phase-flip error rate $e_1^{ph}$ of the single photon state $\ket{01}$ or $\ket{10}$ that Alice decides sending and Bob decides not sending or Alice decides not sending and Bob decides sending in the $Z$ windows. The asymptotic case is considered in Ref.~\cite{wang2018twin}, and there are infinite intensities in $\widetilde{X}$ windows and infinite pulses in the whole protocol, thus $s_1$ and $e_1^{ph}$ could be estimated exactly. 

Alice and Bob send their prepared pulses to Charlie, and Charlie is assumed to perform interferometric measurements on the received pulses and announces the measurement result to Alice and Bob. If one and only one detector clicks in the measurement process, Charlie will also announce whether the left detector or right detector clicks. The effective events of $Z$ windows and $\widetilde{X}$ windows are defined individually: it is an effective event of $Z$ windows if one and only one detector clicks; it is an effective event of $\widetilde{X}$ windows if one and only one detector clicks and Alice and Bob send the coherent state with the same intensity, and their phases satisfy the post-selection criterion, which is 
\begin{equation}\label{crit}
1-|\cos{(\theta_A-\theta_B-\psi_{AB})}|\le|\lambda|,
\end{equation}
where $\theta_A$ and $\theta_B$ are the phases of coherent states prepared by Alice and Bob respectively, and $\psi_{AB}$ can take an arbitrary value which can be different from time to time as Alice and Bob like, so as to obtain a satisfactory key rate for the protocol~\cite{liu2019experimental}. Note that $1-|\cos{[\theta_A-\theta_B-(\gamma_A-\gamma_B)]}|\le|\lambda|$ according to the security proof of Ref.~\cite{wang2018twin} in the poset-selction criterion there~\cite{wang2018twin}, both $\gamma_A$ and $\gamma_B$ can take arbitrary values there~\cite{wang2018twin}. However, in applying the criterion, we only need the value $\gamma_A-\gamma_B$ which is actually only \textit{one} value. Thus we could just use $\psi_{AB}$ in Eq.~\eqref{crit} here. The value of $\lambda$ is decided by the size of phase slice, $\Delta$, that Alice and Bob choose~\cite{lu2018overcoming}. The Eq.~\eqref{crit} is equivalent to 
\begin{equation}
|\theta_A-\theta_B-\psi_{AB}|\le \frac{\Delta}{2},\quad |\theta_A-\theta_B-\psi_{AB}-\pi|\le \frac{\Delta}{2}.
\end{equation}
Same with that in Ref.~\cite{yu2019sending}, here $|x|$ means the degree of the minor angle enclosed by the two rays that enclose the rotational angle of degree $x$, e.g., $|-15\pi/8|=|15\pi/8| = \pi/8$, $|-\pi/10|= \pi/10$. 

The phases of coherent states in $Z$ windows are never be announced in the public channel, thus the coherent states in $Z$ windows are phase-randomized coherent states which are equivalent to classical mixture of different photon numbers. Only the effective events of single-photon states in those $Z$ windows that Alice decides sending and Bob decides not sending or Alice decides not sending and Bob decides sending, are \textbf{untagged} events, thus we have the following formula of final key rate
\begin{equation}
R=2p(1-p)\mu_ze^{-\mu_z}s_1[1-h(e_1^{ph})]-fS_zh(E_z),
\end{equation} 
where $S_z$ is the counting rate of pulses in $Z$ windows and $E_z$ is the corresponding error rate, $h(x)=-x\log_2{x}-(1-x)\log_2{(1-x)}$ is the binary Shannon entropy function, $f$ is the error correction inefficiency, and $s_1$ and $e_1^{ph}$ are defined in the beginning of this section.

However, the number of pulses is finite in practice and thus there can not be infinite intensities in $\widetilde{X}$ windows. Here we consider the four-intensity decoy state SNS protocol~\cite{yu2019sending}. In each time, Alice and Bob randomly choose the decoy window or signal window with probabilities $1-p_z$ and $p_z$. If the decoy window is chosen, Alice (Bob) randomly chooses vacuum state $\ket{0}$, $\ket{e^{i\delta_A}\sqrt{\mu_1}}$ or $\ket{e^{i\delta_A^\prime}\sqrt{\mu_2}}$ (vacuum state $\ket{0}$, $\ket{e^{i\delta_B}\sqrt{\mu_1}}$ or $\ket{e^{i\delta_B^\prime}\sqrt{\mu_2}}$) with probabilities $p_0$, $p_1$ and $1-p_0-p_1$ respectively, where $\delta$ is random in $[0,2\pi)$. If the signal window is chosen, Alice (Bob) randomly chooses vacuum state $\ket{0}$, or phase-randomized weak coherent state of intensity $\mu_z$, with probabilities $p_{z0}$ and $1-p_{z0}$. Then Alice and Bob prepare the chosen states and send them to Charlie. Charlie is assumed to perform interferometric measurements on the received quantum signals and announces the measurement result to Alice and Bob. If one and only one detector clicks in the measurement process, Charlie will also announces whether the left detector or right detector clicks. Then Alice and Bob will take it as an one-detector heralded event. After Alice and Bob repeat the above steps for $N$ times, they perform the following data post-processing steps. 

\noindent 1. \textbf{Sifting.}  If both Alice and Bob choose the signal window, it is a $Z$ window. If both Alice and Bob choose the decoy window, it is an $\widetilde{X}$ window. Besides, we define that if both Alice and Bob decide to send the phase-randomized coherent state with intensity $\mu_1$, as $X_1$ window, which is a subset of $\widetilde{X}$ windows. According to the criterion introduced in the beginning of this section, Alice and Bob decide whether an one-detector heralded event is an effective event. We define three kinds of sets, $\mathcal{Z},\mathcal{X}_1$, and $\mathcal{X}_2$. The set $\mathcal{Z}$ includes all effective events in $Z$ windows. The set $\mathcal{X}_1$ includes all effective events in $X_1$ windows. And the set $\mathcal{X}_2$ includes all other one-detector heralded events.

\noindent 2. \textbf{Parameter estimation.} For the events in the set $\mathcal{Z}$, Alice will denote it as bit $0$ if she sends a vacuum state, and denote it as bit $1$ if she sends a phased-randomized weak coherent state. In the same time, Bob will denote it as bit $1$ if he sends a vacuum state, and denote it as bit $0$ if he sends a phased-randomized weak coherent state. Finally Alice and Bob form the $n_t$-bit strings $Z_s$ and $Z_s^\prime$ according to the events in set $\mathcal{Z}$. Then through the decoy-state method, Alice and Bob estimate $n_1$ according to the events in $\mathcal{X}_2$ and estimate $e_1^{ph}$ according the events in set $\mathcal{X}_1$, where $n_1$ is the lower bound of bits caused by \textbf{untagged} events in $Z_s$ or $Z_s^\prime$, and 
$e_1^{ph}$ is the upper bound of phase-flip error rate of the \textbf{untagged} bits. The details of how to calculate $n_1$ and $e_1^{ph}$ are shown in the Methods part.

\noindent 3. \textbf{Error correction.} Alice and Bob perform an information reconciliation scheme to correct $Z_s^\prime$, and Bob will obtain an estimate $\hat{Z_s}$ of $Z_s$ from $Z_s^\prime$. To achieve this goal, Alice sends Bob $leak_{EC}$ bits of error correction data. Then Alice computes a hash of $Z_s$ of length $\log_2{(1/\varepsilon_{cor})}$ using a random universal hash function, and she sends the hash and hash function to Bob~\cite{renner2005security}. If the hash that Bob computes is the same with Alice, the probability that $Z_s$ and $\hat{Z_s}$ aren't the same, Pr$({Z_s \neq \hat{Z_s}})$, is less than $\varepsilon_{cor}$. If the hash that Bob computes is not the same with Alice, the protocol aborts.

\noindent 4. \textbf{Private amplification.} Alice and Bob apply a privacy amplification scheme based on two-universal hashing~\cite{renner2005security} to extract two shorter strings of length $l$ from $Z_s$ and $\hat{Z_s}$. Alice and Bob obtain strings $Z_{PA}$ and $\hat{Z}_{PA}$ which is the final secret key after privacy amplification.

The protocol is $\varepsilon_{cor}$-correct if the error correction step is passed. If the final length of secret keys, $l$, satisfies
\begin{equation}\label{rr}
\begin{split}
l=&n_1[1-h(e_1^{ph})]-leak_{EC}-\log_2{\frac{2}{\varepsilon_{cor}}}\\
&-2\log_2{\frac{1}{\sqrt{2}\varepsilon_{PA}\hat{\varepsilon}}},
\end{split}
\end{equation}
the  protocol is $\varepsilon_{sec}$-secret. And according to the composable framework, the security coefficient of the whole protocol is $\varepsilon_{tol}=\varepsilon_{cor}+\varepsilon_{sec}$, where 
$\varepsilon_{sec}=2\hat{\varepsilon}+4\bar{\varepsilon}+\varepsilon_{PA}+\varepsilon_{n_1}$. Here, $\varepsilon_{cor}$ is the failure probability of error correction; $\bar{\varepsilon}$ is the accuracy of estimating the smooth min-entropy, which is also the failure probability that the real value of $e_1^{ph}$ isn't in the bound that we estimate; $\varepsilon_{PA}$ is the failure probability of privacy amplification; $\varepsilon_{n_1}$ is the failure probability that the real value of $n_1$ isn't in the bound that we estimate. The value of $leak_{EC}$ is related to the specific error correction schemes, and in general $leak_{EC}=fn_th(E_z)$, where $E_z$ is the error rate of strings $Z_s$ and $Z_s^\prime$. 

\section{Numerical simulation}\label{simulation}
If an experiment of SNS protocol is done, we can first calculate the lower and upper bound of $\mean{S_{jk}}$ with Eqs.~\eqref{eq10}, \eqref{mul}-\eqref{sjklower} from their observed values. And we can get the upper bound of $\mean{T_{\Delta}}$ in a similar way. Then we can get the lower bound of $\mean{s_1^Z}$ and the upper bound of $\mean{e_1^{ph}}$ with Eqs.~\eqref{s11l} and \eqref{e11u}. Then we can get the lower bound of $n_1$ and the upper bound of $e_1^{ph}$ with Eqs~\eqref{observ}-\eqref{n1e1}. Finally, we can get how many bits of secret keys we could extract from this experiment with Eq.~\eqref{rr}. The problem is that we do not have such observed values and we need to simulate what values we would observe in the experiment with the experimental parameters list in Table.~\ref{exproperty}. All symbols appearing in this paragraph is defined in Sec.~\ref{calculation}.
\begin{table}%\footnotesize
\begin{ruledtabular}
\begin{tabular}{ccccccc}
$p_d$&$e_0$ & $e_d$ &$\eta_d$ & $f$ & $\alpha_f$&$\xi$ \\%%\rule{0pt}{0.3cm}\\
\hline
$1.0\times10^{-10}$&0.5& $15\%$  & $80.0\%$ & $1.1$ & $0.2$&$1.0\times 10^{-10}$ \\ %%\rule{0pt}{0.3cm}\\
%%\hline
%%\hline
\end{tabular}
\end{ruledtabular}
\caption{List of experimental parameters used in numerical simulations. Here $p_d$: the dark count rate of Charlie's detectors; $e_0$: error rate of the vacuum count; $e_d$: the misalignment-error probability; $\eta_d$: the detection efficiency of Charlie's detectors; $f$: the error correction inefficiency; $\alpha_f$: the fiber loss coefficient ($dB/km$); $\xi$: the failure probability of statistical fluctuation analysis.}\label{exproperty}
\end{table}

We use the linear model to simulate the observed values of experiment with the experimental parameters list in Table.~\ref{exproperty}. Without loss of generality, we assume the distance between Alice and Charlie and the distance between Bob and Charlie are the same, and we assume the property of Charlie's two detectors are the same. The total transmittance of the experiment set-ups is $\eta=10^{-L/100}\eta_d$, where $L$ is the distance between Alice and Bob. The simulation of those observed values are shown in Sec.~\ref{simu}, which are related to $\eta$ and other parameters list in Table.~\ref{exproperty}.

Here we set $\varepsilon_{cor}=\hat{\varepsilon}=\varepsilon_{PA}=\xi,\bar{\varepsilon}=4\xi$ and $\varepsilon_{n_1}=4\xi$, and thus security coefficient of the whole protocol is $\varepsilon_{tol}=24\xi=2.4\times 10^{-9}$. The reason we set $\bar{\varepsilon}=4\xi$ and $\varepsilon_{n_1}=4\xi$ is that we use the Chernoff bound for four times to estimate $e_1^{ph}$ and $n_1$ (Notice that we could handle $\mean{\underline{S}_{01}},\mean{\underline{S}_{10}}$ and $\mean{\overline{S}_{02}},\mean{\overline{S}_{20}}$ together in Eq.~\eqref{eq10}). In order to fairly compare the performance of generating final keys of different total pulse numbers, $N$, we define the key rate of per sending pulse, $R=l/N$.

\begin{figure}
\centering
\includegraphics[width=8cm]{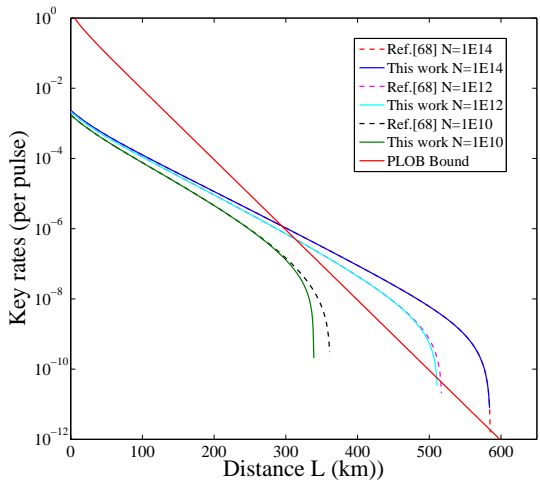}
\caption{The optimal key rates (per pulse) versus transmission distance (the distance between Alice and Bob) with the results of this work and Ref.~\cite{yu2019sending} under the experimental parameters listed in Table \ref{exproperty}. The dashed lines are results of Ref.~\cite{yu2019sending} and the solid lines are the results of this work. Here we simulate three groups of results where $N=1\times10^{14},1\times10^{12},1\times10^{10}$. Here the red solid line is the PLOB bound.}\label{figure1}
\end{figure}

\begin{figure}
\centering
\includegraphics[width=8cm]{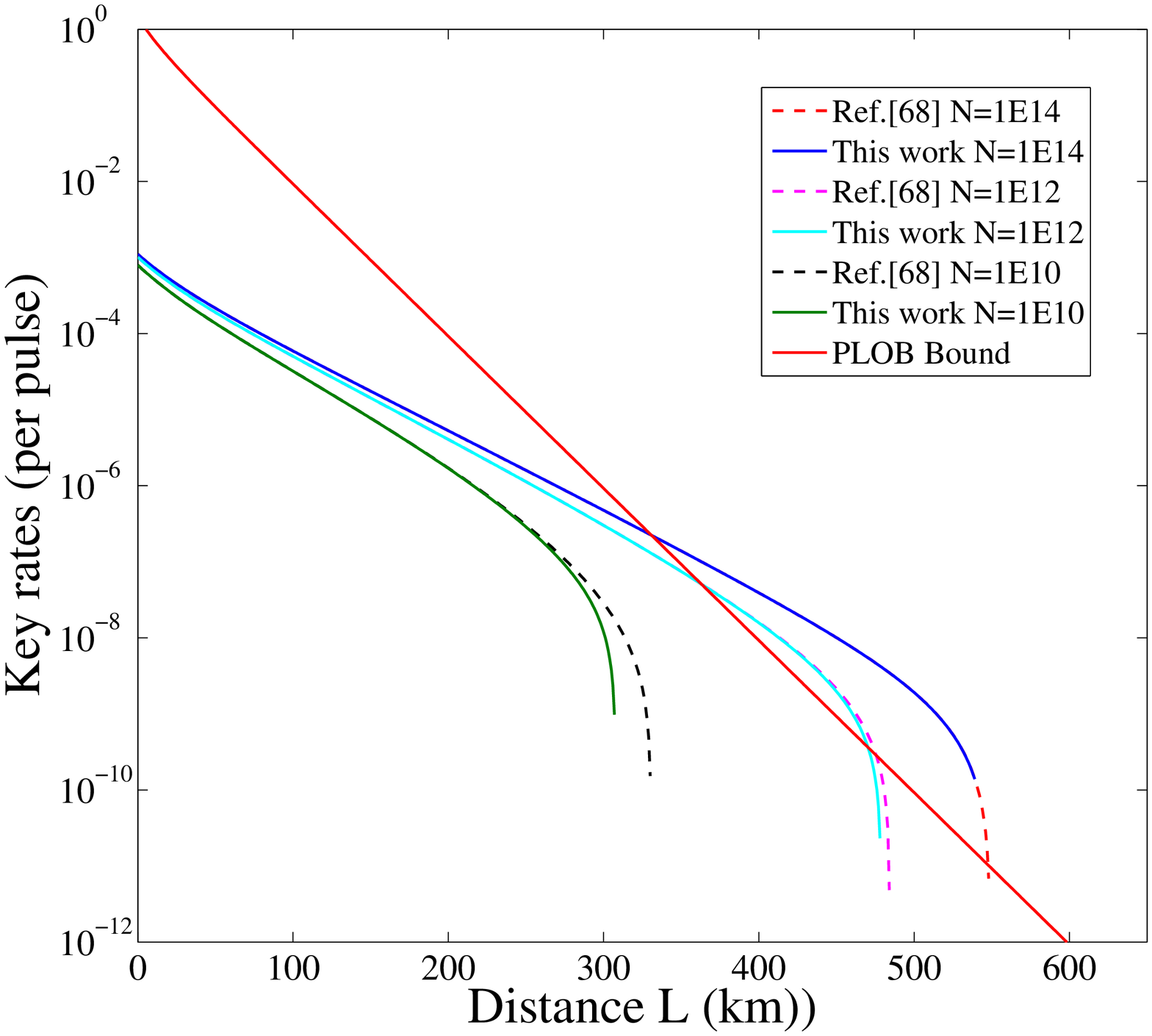}
\caption{The optimal key rates (per pulse) versus transmission distance (the distance between Alice and Bob) with the results of this work and Ref.~\cite{yu2019sending} under the experimental parameters listed in Table \ref{exproperty}, except we set $e_d=20\%$. The dashed lines are results of Ref.~\cite{yu2019sending} and the solid lines are the results of this work. Here we simulate three groups of results where $N=1\times10^{14},1\times10^{12},1\times10^{10}$. Here the red solid line is the PLOB bound.}\label{figure2}
\end{figure}

\begin{figure}
\centering
\includegraphics[width=8cm]{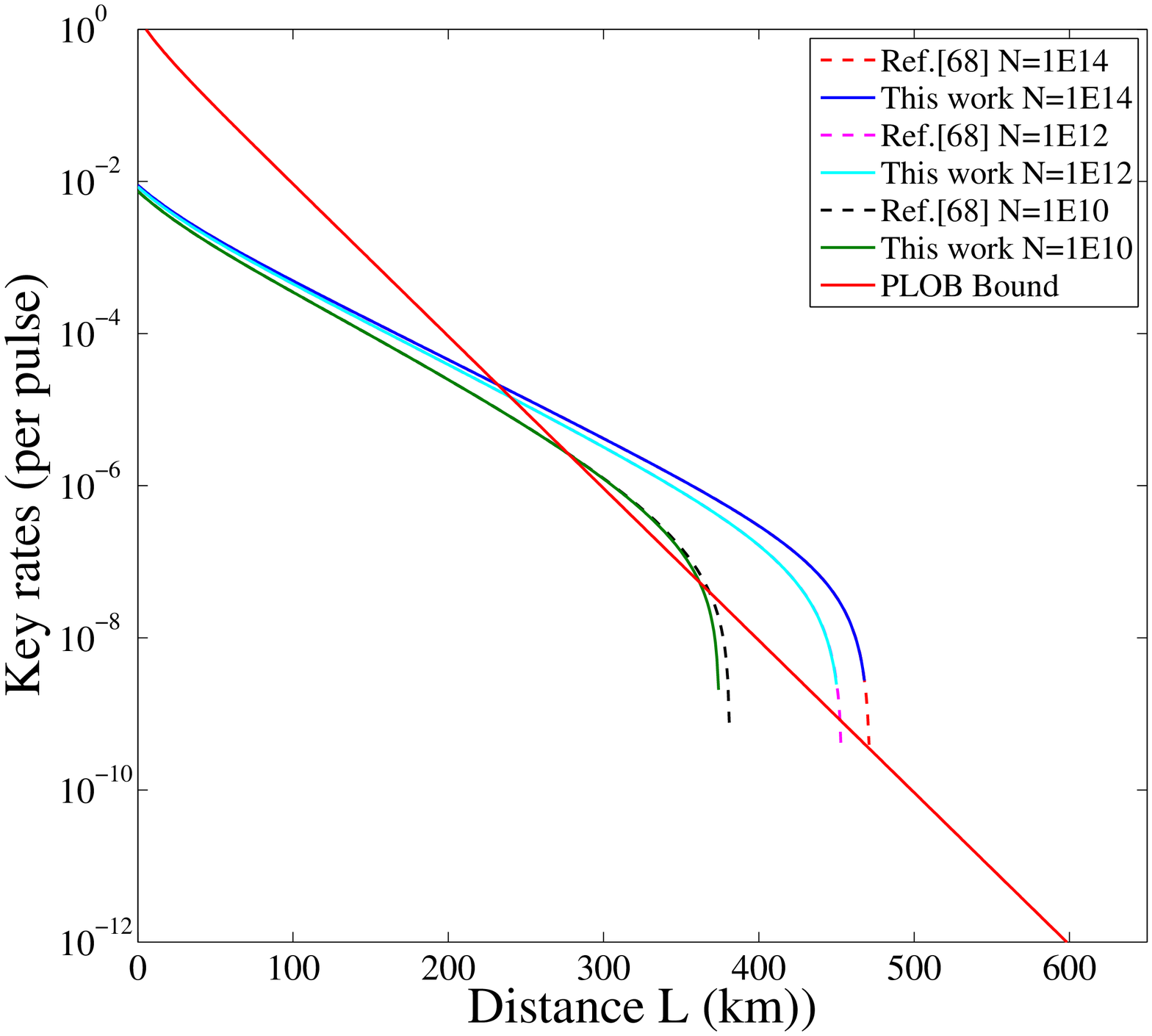}
\caption{The optimal key rates (per pulse) versus transmission distance (the distance between Alice and Bob) with the results of this work and Ref.~\cite{yu2019sending}. Here we set $p_d=1\times10^{-8}$ and $e_d=5\%$, the other experimental parameters we use are listed in Table \ref{exproperty}. The dashed lines are results of Ref.~\cite{yu2019sending} and the solid lines are the results of this work. Here we simulate three groups of results where $N=1\times10^{14},1\times10^{12},1\times10^{10}$. Here the red solid line is the PLOB bound.}\label{figure3}
\end{figure}

\begin{figure}
\centering
\includegraphics[width=8cm]{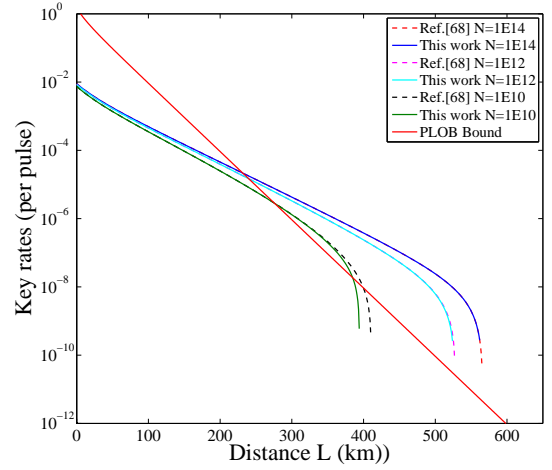}
\caption{The optimal key rates (per pulse) versus transmission distance (the distance between Alice and Bob) with the results of this work and Ref.~\cite{yu2019sending}. Here we set $p_d=1\times10^{-9}$ and $e_d=5\%$, the other experimental parameters we use are listed in Table \ref{exproperty}. The dashed lines are results of Ref.~\cite{yu2019sending} and the solid lines are the results of this work. Here we simulate three groups of results where $N=1\times10^{14},1\times10^{12},1\times10^{10}$. Here the red solid line is the PLOB bound.}\label{figure4}
\end{figure}

Fig.~\ref{figure1} and Fig.~\ref{figure2} are our simulation results of this work and Ref.~\cite{yu2019sending} with the experimental parameters list in Table.~\ref{exproperty}. The only difference of Fig.~\ref{figure1} and Fig.~\ref{figure2} is that $e_d=15\%$ in Fig.~\ref{figure1} and $e_d=20\%$ in Fig.~\ref{figure2}. The results of this work and Ref.~\cite{yu2019sending} is almost overlap while we set $N=1\times10^{14}$, but the difference of the results is obvious while we set $N=1\times 10^{10}$, especially in the end of the lines. Still, the secure distance of SNS protocol can still reach up to 500 $km$ with $20\%$ misalignment error and $1\times 10^{12}$ total pulses, even if we take all the effects of finite-key size into consideration.

Fig.~\ref{figure3} and Fig.~\ref{figure4} are our simulation results of another two groups of experimental parameters. We set $p_d=1\times10^{-8}$ and $e_d=5\%$ in Fig.~\ref{figure3} and $p_d=1\times10^{-9}$ and $e_d=5\%$ in Fig.~\ref{figure4}. The other experimental parameters we use are listed in Table \ref{exproperty}. Same with Fig.~\ref{figure1} and Fig.~\ref{figure2}, we simulate three groups of results where $N=1\times10^{14},1\times10^{12},1\times10^{10}$. Comparing Fig.~\ref{figure3} with Fig.~\ref{figure4}, we can find that the secure distances are improved at most 100 $km$ if the dark count is reduced by an order of magnitude. Still, the complete effect of finite size is reflected in the end of lines especially when the total number of pulses, $N$, is relatively small.

\section{Conclusion}
In this paper, we show an analysis of the finite-key size effect of SNS protocol and get the relation of final key length $l$ and the security coefficient, as shown in Eq.~\eqref{rr}. Eq.~\eqref{rr} is derived by the method proposed in Ref.~\cite{tomamichel2012tight}, and thus it can produce unconditional secure final key under general attack, including all coherent attacks. The numerical results show that the secure distance of SNS protocol can still reach up to 500 $km$ with $20\%$ misalignment error and $1\times10^{12}$ total pulses, even if we take all the effects of finite-key size into consideration. This clearly shows that the SNS protocol~\cite{wang2018twin} of TF-QKD is on the one hand secure under general attack, i.e., as secure as the existing decoy-state MDI-QKD, on the other hand more efficient than the existing decoy-state MDI-QKD by many orders of magnitudes in key rate at long distance domain.

{\bf{Acknowledgement:}} We thank Hai Xu for discussions. We acknowledge the financial support in part by Ministration of Science and Technology of China through The National Key Research and Development Program of China grant No. 2017YFA0303901; National Natural Science Foundation of China grant No. 11474182, 11774198 and U1738142.

\section*{Methods}
\subsection{The relation of the length of final key and $\varepsilon_{sec}$}
In this protocol, any attack to quantum channel and detectors is allowed only if it doesn't break the rules of quantum mechanics, and we call the attacker as Eve. We denote the system of Eve after error correction as $E^\prime$. If Alice and Bob apply a privacy amplification scheme based on two-universal hashing to extract two shorter strings of length $l$ from $Z_s$, the protocol is $\varepsilon_{sec}$-secret~\cite{renner2005security,tomamichel2010duality}
\begin{equation}\label{sec}
\varepsilon_{sec}\le 2\varepsilon+\frac{1}{2}\sqrt{2^{l-H_{min}^\varepsilon(Z_s|E^\prime)}},
\end{equation}  
where $H_{min}^\varepsilon(Z_s|E^\prime)$ is the $\varepsilon$-smooth min entropy. It measures the max probability of guessing $Z_s$ right giving $E^\prime$. $E^\prime$ could be decomposed as $CE$, where $C$ is the system of leakage information while Alice and Bob perform error correction and $E$ is the system of Eve before error correction. According to the chain rules~\cite{renner2005security}, we have
\begin{equation}
H_{min}^\varepsilon(Z_s|E^\prime)\ge H_{min}^\varepsilon(Z_s|E)-|C|,
\end{equation}
where $|C|<leak_{EC}+\log_2(\frac{2}{\varepsilon_{cor}})$. And we could decompose the string $Z_s$ as $Z_1Z_{rest}$, where $Z_1$ is the bits caused by \textbf{untagged}-photon events and $Z_{rest}$ is the other bits of $Z_s$~\cite{curty2014finite}. Thus according to the chain rules~\cite{vitanov2013chain}, we have
\begin{equation}
\begin{split}
H_{min}^\varepsilon(Z_s|E)\ge& H_{min}^{\bar{\varepsilon}}(Z_1|Z_{rest}E)+H_{min}^{\varepsilon^\prime}(Z_{rest}|E)\\
&-2\log_2{\frac{\sqrt{2}}{\hat{\varepsilon}}},
\end{split}
\end{equation} 
where $\varepsilon=2\bar{\varepsilon}+\varepsilon^\prime+\hat{\varepsilon}$ and $H_{min}^{\varepsilon^\prime}(Z_{rest}|E)\ge0$.

Besides, we denote $X$ basis as $\{\frac{1}{2}(\ket{01}+e^{i\theta}\ket{10}),\frac{1}{2}(\ket{01}-e^{i\theta}\ket{10})\}$ and $Z$ basis as $\{\ket{01},\ket{10}\}$, where $\theta$ can be an arbitary value. To get the lower bound of $H_{min}^{\bar{\varepsilon}}(Z_1|Z_{rest}E)$, we need to use the uncertainty relation of smooth min and max entropy~\cite{tomamichel2012tight,tomamichel2011uncertainty}. It says that if the \textbf{untagged}-photon states prepared in $X$ basis and $Z$ basis are orthogonal unbiased, and if the states originally prepared and measured under the Z-basis are now prepared and measured under the X-basis and obtained strings $X_{s1}$ and $X_{s1}^\prime$ by Alice and Bob respectively, then we have
\begin{equation}
\begin{split}
H_{min}^{\bar{\varepsilon}}(Z_1|Z_{rest}E)&\ge n_1- H_{max}^{\bar{\varepsilon}}(X_{s1}|X_{s1}^\prime)\\
&\ge n_1-n_1h(e_1^{ph}).
\end{split}
\end{equation}
Finally we have
\begin{equation}\label{hhmin}
\begin{split}
H_{min}^\varepsilon(Z_s|E^\prime)\ge & n_1[1-h(e_1^{ph})]-leak_{EC}\\
& -\log_2{\frac{2}{\varepsilon_{cor}}}-2\log_2{\frac{\sqrt{2}}{\hat{\varepsilon}}}.
\end{split}
\end{equation}
Combining Eqs.~\eqref{rr},~\eqref{sec} and~\eqref{hhmin} and setting $\varepsilon^\prime=0$, we have 
\begin{equation}
\varepsilon_{sec}\le 2\hat{\varepsilon}+4\bar{\varepsilon}+\varepsilon_{PA}.
\end{equation}
Finally, containing the failure probability that the real value of $n_1$ isn't in the bound that we estimate, $\varepsilon_{n_1}$, we have 
\begin{equation}
\varepsilon_{sec}\le 2\hat{\varepsilon}+4\bar{\varepsilon}+\varepsilon_{PA}+\varepsilon_{n_1}.
\end{equation}
\subsection{The calculation method of $n_1$ and $e_1^{ph}$}\label{calculation}
The method we use here is similar with Ref. \cite{yu2019sending}. In an $\widetilde{X}$ window with different intensities from Alice and Bob, they don't announce any phase information in the protocol, therefore the coherent states sent out from each sides can be regarded as classical mixture of different photon numbers. We denote $\rho=\oprod{0}{0}, \rho_1=\sum_{k=0}\frac{\mu_1^ke^{-\mu_1}}{k!}\oprod{k}{k}, \rho_2=\sum_{k=0}\frac{\mu_2^ke^{-\mu_2}}{k!}\oprod{k}{k}$ and $\rho_z=\sum_{k=0}\frac{\mu_z^k}{k!}\oprod{k}{k}$, where $\rho_1$ and $\rho_2$ are the density operator of the coherent states used here in $\widetilde{X}$ windows. And this also applies to Bob's quantum state. In the whole protocol, Alice and Bob obtain $N_{jk}(jk=\{00,01,02,10,20\})$ instances when Alice sends state $\rho_j$ and Bob sends state $\rho_k$. And after the sifted step, Alice and Bob obtain $n_{jk}$ one-detector heralded events. We denote the counting rate of source $jk$ as $S_{jk}=n_{jk}/N_{jk}$. With all those definitions, we have 
\begin{equation}\label{eq10}
\begin{split}
N_{00}=&[(1-p_z)^2p_0^2+2(1-p_z)p_zp_0p_{z0}]N,\\
N_{01}=&N_{10}=[(1-p_z)^2p_0p_1+(1-p_z)p_zp_{z0}p_1]N,\\
N_{02}=&N_{20}=[(1-p_z)^2(1-p_0-p_1)p_0\\
&+(1-p_z)p_zp_{z0}(1-p_0-p_1)]N.
\end{split}
\end{equation}

Besides, we need define two new subsets of $X_1$ windows, $C_{\Delta^+}$ and $C_{\Delta^-}$, to estimate the upper bound of $e_1^{ph}$. $C_{\Delta^+}$ contains all the instance that both Alice and Bob prepare $\ket{e^{i\delta_A}\sqrt{\mu_1}}$ and $\ket{e^{i\delta_B}\sqrt{\mu_1}}$ and  
$|\delta_A-\delta_B|\le \frac{\Delta}{2}$. $C_{\Delta^-}$ contains all the instance that both Alice and Bob prepare $\ket{e^{i\delta_A}\sqrt{\mu_1}}$ and $\ket{e^{i\delta_B}\sqrt{\mu_1}}$ and $|\delta_A-\delta_B-\pi|\le \frac{\Delta}{2}$. Same with that in Ref.~\cite{yu2019sending}, here $|x|$ means the degree of the minor angle enclosed by the two rays that enclose the rotational angle of degree $x$, e.g., $|-15\pi/8|=|15\pi/8| = \pi/8$, $|-\pi/10|= \pi/10$. The number of instances in $C_{\Delta^\pm}$ is
\begin{equation}\label{eq11}
N_{\Delta^\pm}=\frac{\Delta}{2\pi}(1-p_z)^2p_1^2N.
\end{equation}

We denote the number of effective events of right detectors responding from $C_{\Delta^+}$ as $n_{\Delta^+}^R$, and the number of effective events of left detectors responding from $C_{\Delta^-}$ as $n_{\Delta^-}^L$. And we get the counting error rate of $C_{\Delta^{\pm}}$, $T_\Delta=\frac{n_{\Delta^+}^R+n_{\Delta^-}^L}{2N_{\Delta^{\pm}}}$.

If we denote the expected value of the counting rate of \textbf{untagged} photons as $\mean{s_1^Z}$, the lower bound of $\mean{s_1^Z}$ is 
\begin{equation}\label{s11l}
\begin{split}
\mean{s_1^Z}&\ge \mean{\underline{s}_1^Z}=\frac{1}{2\mu_1\mu_2(\mu_2-\mu_1)}[\mu_2^2e^{\mu_1}(\mean{\underline{S}_{01}}+\mean{\underline{S}_{10}})\\
&-\mu_1^2e^{\mu_2}(\mean{\overline{S}_{02}}+\mean{\overline{S}_{20}})-2(\mu_2^2-\mu_1^2)\mean{\overline{S}_{00}}],
\end{split}
\end{equation}
where $\mean{S_{jk}}$ is the expected value of $S_{jk}$, and $\mean{\overline{S}_{jk}}$ and $\mean{\underline{S}_{jk}}$ are the upper bound and lower bound of $\mean{S_{jk}}$ when we estimate the expected value from its observed value.

The expected value of the phase-flip error rate of the \textbf{untagged} photons satisfies~\cite{yu2019sending}
\begin{equation}\label{e11u}
\mean{e_1^{ph}}\le \mean{\overline{e}_1^{ph}}=\frac{\mean{\overline{T}_\Delta}-\frac{1}{2}e^{-2\mu_1}\mean{\underline{S}_{00}}}{2\mu_1e^{-2\mu_1}\mean{\underline{s}_1^Z}}.
\end{equation}
Here we use the fact that the error rate of vacuum state is always $\frac{1}{2}$.

\noindent \textbf{Chernoff bound}. The formulas of $\mean{\underline{s}_1^Z}$ and $\mean{\overline{e}_1^{ph}}$ are represented by expected values, but the values we get in experiment are observed values. To close the gap between the expected values and observed values, we need Chernoff bound~\cite{jiang2017measurement,chernoff1952measure}. Let $X_1,X_2,\dots,X_n$ be $n$ random samples, detected with the value 1 or 0, and let $X$ denote their sum satisfying $X=\sum_{i=1}^nX_i$. $\phi$ is the expected value of $X$. We have
\begin{align}
\label{mul}\phi^L(X)=&\frac{X}{1+\delta_1(X)},\\
\label{muu}\phi^U(X)=&\frac{X}{1-\delta_2(X)},
\end{align}
where we can obtain the values of $\delta_1(X)$ and $\delta_2(X)$ by solving the following equations
\begin{align}
\label{delta1}\left(\frac{e^{\delta_1}}{(1+\delta_1)^{1+\delta_1}}\right)^{\frac{X}{1+\delta_1}}&=\frac{\xi}{2},\\
\label{delta2}\left(\frac{e^{-\delta_2}}{(1-\delta_2)^{1-\delta_2}}\right)^{\frac{X}{1-\delta_2}}&=\frac{\xi}{2},
\end{align}
where $\xi$ is the failure probability. Thus we have 
\begin{equation}\label{sjklower}
\phi^L({N_{jk}S_{jk}})=N_{jk}\mean{\underline{S}_{jk}},\phi^U({N_{jk}S_{jk}})=N_{jk}\mean{\overline{S}_{jk}}.
\end{equation}
Still Eqs.~\eqref{s11l} and \eqref{e11u} are the lower bound of the expected values of the counting rate and the upper bound of the phase flip error rate of single-photons. The final question is what their real values are in this specific experiment, and we need the Chernoff bound to help us estimate their real values from their expected values. Similar to Eqs.~\eqref{mul}- \eqref{delta2}, the observed value, $\varphi$, and its expected value, $Y$, satisfy
\begin{align}
\label{observ}&\varphi^U(Y)=[1+\delta_1^\prime(Y)]Y,\\
&\varphi^L(Y)=[1-\delta_2^\prime(Y)]Y,
\end{align}   
where we can obtain the values of $\delta_1^\prime(Y)$ and $\delta_2^\prime(Y)$ by solving the following equations
\begin{align}
\left(\frac{e^{\delta_1^\prime}}{(1+\delta_1^\prime)^{1+\delta_1^\prime}}\right)^{Y}&=\frac{\xi}{2},\\
\left(\frac{e^{-\delta_2^\prime}}{(1-\delta_2^\prime)^{1-\delta_2^\prime}}\right)^{Y}&=\frac{\xi}{2}.
\end{align}
We define $N_1=2p_z^2p_{z0}(1-p_{z0})\mu_ze^{-\mu_z}N$, and we have~\cite{yu2019sending}
\begin{equation}\label{n1e1}
n_1=\varphi^L(N_1\mean{\underline{s}_1^Z}),\quad e_1^{ph}=\frac{\varphi^U(N_1\mean{\underline{s}_1^Z}\mean{\overline{e}_1^{ph}})}{N_1\mean{\underline{s}_1^Z}}.
\end{equation} 
This ends the estimate of $n_1$ and $e_1^{ph}$.
 
\subsection{The simulation of observed values}\label{simu}
We use the linear model to simulate the observed values of experiment with the experimental parameters list in Table.~\ref{exproperty}. Without loss of generality, we assume the distance between Alice and Charlie and the distance between Bob and Charlie are the same, and we assume the properties of Charlie's two detectors are the same. If the total transmittance of the experiment set-ups is $\eta$, then we have
\begin{align*}
n_{00}&=2p_d(1-p_d)N_{00},\\
n_{01}&=n_{10}=2[(1-p_d)e^{\eta\mu_1/2}-(1-p_d)^2e^{-\eta\mu_1}]N_{01},\\
n_{02}&=n_{20}=2[(1-p_d)e^{\eta\mu_2/2}-(1-p_d)^2e^{-\eta\mu_2}]N_{02},\\
n_t&=n_{signal}+n_{error},\\
E_z&=\frac{n_{error}}{n_t},\\
n_{\Delta^+}^R&=n_{\Delta^-}^L=[T_X(1-2e_d)+e_dS_X]N_{\Delta^{\pm}},
\end{align*}

where $N_{00},N_{01},N_{10},N_{02},N_{20},N_{\Delta^{\pm}}$ are defined in Eqs.~\eqref{eq10} and \eqref{eq11} and
\begin{align*}
n_{signal}=&4Np_z^2p_{z0}(1-p_{z0})[(1-p_d)e^{-\eta\mu_z/2}\\
&-(1-p_d)^2e^{-2\eta\mu_z}],\\
n_{error}=&2Np_z^2(1-p_{z0})^2[(1-p_d)e^{-\eta\mu_z}I_0(\eta\mu_z)\\
&-(1-p_d)^2e^{-2\eta\mu_z}]+2Np_z^2p_{z0}^2p_d(1-p_d),\\
T_X=&\frac{1}{\Delta}\int_{-\frac{\Delta}{2}}^{\frac{\Delta}{2}}(1-p_d)e^{-2\eta\mu_1\cos^2{\frac{\delta}{2}}}d\delta\\
&-(1-p_d)^2e^{-2\eta\mu_1},\\
S_X=&\frac{1}{\Delta}\int_{-\frac{\Delta}{2}}^{\frac{\Delta}{2}}(1-p_d)e^{-2\eta\mu_1\sin^2{\frac{\delta}{2}}}d\delta\\
&-(1-p_d)^2e^{-2\eta\mu_1}+T_X,
\end{align*}
where $I_0(x)$ is the $0$-order hyperbolic Bessel functions of the first kind. 

\bibliography{refs}

\end{document}